\documentclass[preprint,showpacs,preprintnumbers,amsmath,amssymb]{revtex4}
\usepackage{graphicx}% Include figure files
\usepackage{dcolumn}% Align table columns on decimal point
\usepackage{bm}% bold math

\begin{document}

%\preprint{Coherent phonons}

\title{Differential reflection spectroscopy on InAs/GaAs quantum dots}% Force line breaks with \\

\author{E.W. Bogaart}
\email{e.w.bogaart@tue.nl}
%\altaffiliation[Also at ]{eiTT/COBRA Inter-University Research
%Institute, Eindhoven University of Technology//
%Physics Department P.O.BOX 513, 5600 MB Eindhoven, The Netherlands.}%Lines break automatically or can be forced with \\
\author{J.E.M. Haverkort}
\affiliation{Eindhoven University of
Technology\\ Department of Applied Physics P.O. Box 513, 5600 MB Eindhoven, The
Netherlands}
%\homepage{http://www.phys.tue.nl/hgf/}


\begin{abstract}
In this report, we present the derivation of the differential reflection spectrum as has been reported in \emph{Phys. Rev. B} \textbf{72}, 195301 (2005).
\end{abstract}

\maketitle

\begin{figure}\begin{center}
\includegraphics[width=.9\linewidth]{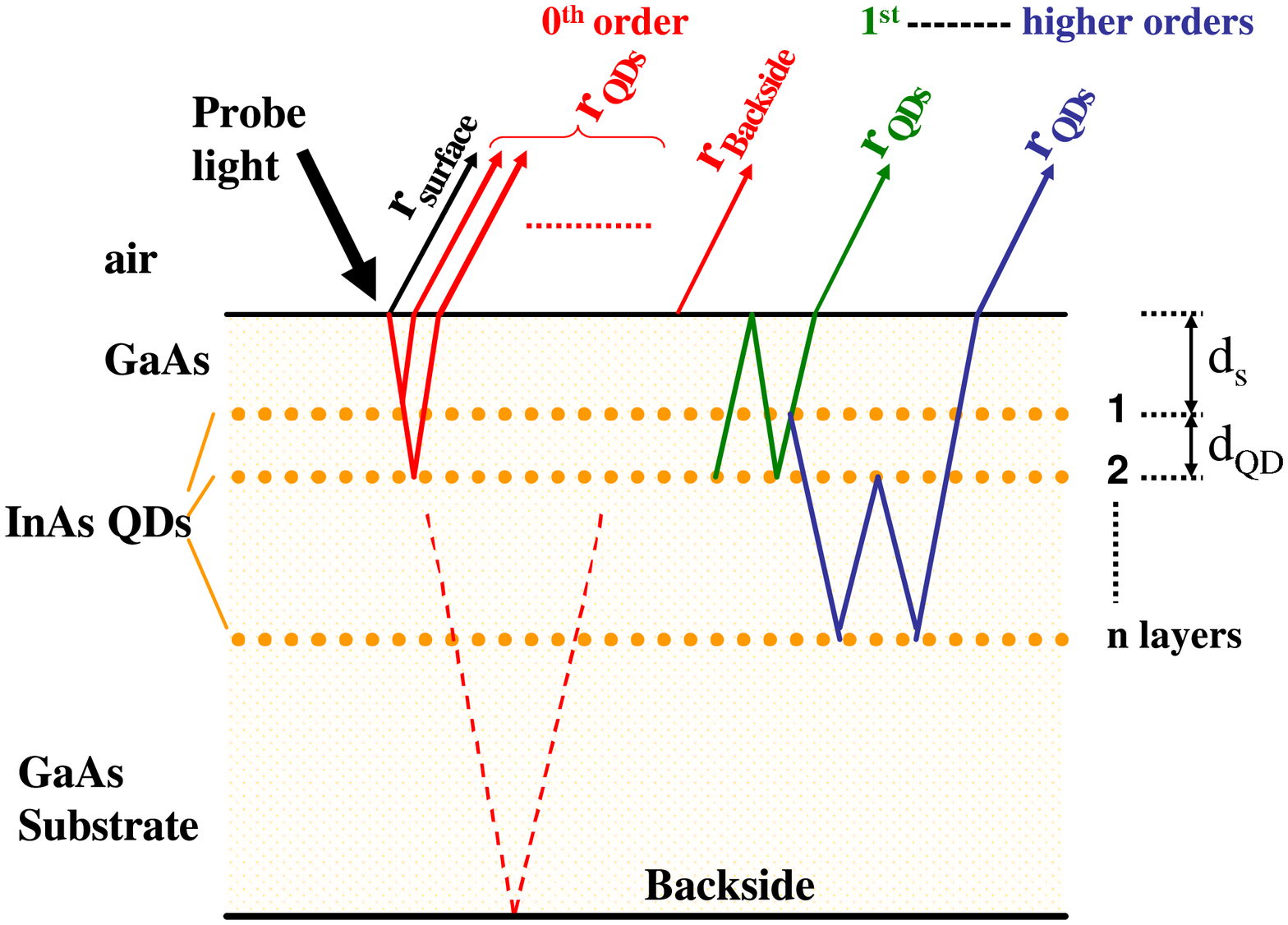}
\caption{Illustration of a QD nanostructure with $n$ layers of QDs. The probe light is partly reflected and partly transmitted at the air-GaAs boundary. Subsequently, the transmitted light will be reflected by the different layers in the structure, and contribute to the total reflection. The order of the reflection is determined by the number of reflections within the structure.}\label{figqdlagen}\end{center}
\end{figure}

%\section{Differential reflection spectroscopy}\label{secdr}
The differential reflection spectrum is derived based on a multiple layer structure with identical QDs, as schematically depicted in Fig. \ref{figqdlagen}. We will consider InAs QDs embedded in GaAs with an optical transition energy of the ground state far below the GaAs band gap energy (1.52 eV at 5 K). This assumption is well justified for InAs QDs, which have ground state transition energies of 1.1 eV or less \cite{bog05a}. The interaction of radiation and free carriers described by the Drude model, which is important at terahertz frequencies, is neglected in our derivations.

For the derivation of the differential reflection equation, a procedure similar to that of Tassone \emph{et al.} \cite{tas92} is used. Although, this method was initially developed to describe the response of quantum well systems, we will assume that the model is also suitable for the description of the reflection response of QD nanostructures. Justification of this assumption lays in the fact that the QD size is small compared to the wavelength of the probe laser. That is, the QDs can be treated as electrically small objects \cite{lak92,sle01}. The electromagnetic response of the nanostructure can be derived by means of an effective medium theory \cite{sle01}. Because the QD volume fraction $f_{QD}$ is significantly smaller than the volume fraction of the host medium, we will use the Maxwell-Garnett approach \cite{max04,lak93,pras04,asp82} to describe the dielectric properties of the nanocomposite. In this approach, the effective permittivity $\varepsilon_{eff}$ of the QD layer is given by
\begin{equation}
\frac{\varepsilon_{eff} -\varepsilon_{h}}{\varepsilon_{eff}+2\varepsilon_{h}} = f_{QD} \frac{\varepsilon_{QD} -\varepsilon_{h}}{\varepsilon_{QD}+2\varepsilon_{h}},\label{eqmaxgar}
\end{equation}
with $\varepsilon_{QD}$ ($\varepsilon_{h}$) the permittivity of the QDs (GaAs host medium). For the calculations as presented below, a QD layer will be approximated by a thin layer with an average refractive index, $n_{QD}=\sqrt{\varepsilon_{eff}}$.

If an electromagnetic wave propagates across an interface between two media with different optical properties, its intensity will be divided between a refracted and a reflected wave. Assume that the wave propagates from vacuum into a dielectric medium with complex refractive index $N$ = $n+i\kappa$, with $n$ the real refractive index and $\kappa$ the attenuation index. The boundary conditions arising from Maxwell's equations require that the tangential components of the electric and magnetic part are continuous across the boundary. Using the boundary conditions, the reflection of the electromagnetic wave can be presented by the Fresnel's formulae \cite{bor99,abe72,ped93}
\begin{eqnarray}
r_n &=&\frac{\cos \theta -(N^2 -\sin^2 \theta)^{1/2}}{\cos\theta +(N^2 -\sin^2 \theta)^{1/2}},\label{eqrn}\\
r_p &=& \frac{N^2 \cos \theta -(N^2 -\sin^2 \theta)^{1/2}}{N^2 \cos \theta +(N^2 -\sin^2 \theta)^{1/2}},\label{eqrp}
\end{eqnarray}
with $r_n$ and $r_p$ the reflection coefficient for the electromagnetic field components perpendicular and parallel to the plane of incidence, respectively, and $\theta$ denotes the angle of incidence. The transmission coefficients are given by $t_n= r_n +1$ and $N t_p= r_p +1$. The attenuation index of the dielectric medium causes a phase shift, and hence, $r_n$ and $r_p$ become complex. In case of normal incidence $\theta$ = 0, the complex reflection coefficients are written by:
\begin{equation}
r_p =-r_n= r= \frac{n+i\kappa -1}{n+i\kappa+1}=|r|e^{i\delta},
\end{equation}
with $\delta$ the phase shift gained by the wave, which is given by $\tan \delta$= $2\kappa/(n^2 +\kappa^2-1)$. Note that also the transmitted wave has a phase shift. The intensity of the transmitted electromagnetic field is given by \cite{abe72}:
\begin{equation}
I_t = I_0\frac{(1-R)^2 e^{-\alpha d} (1-\kappa^2 /n^2)}{1-R^2 e^{-2\alpha d}},\label{eqeint}
\end{equation}
with $I_0$ = $|E_0|^2$ the field intensity before penetration into the dielectric medium. $R$ denotes the reflectivity given by $R$= $|r|^2$, $\alpha$ is the absorption coefficient of the medium, and $d$ is the penetration depth.

A schematic illustration of the QD nanostructure and the reflections induced by each layer is depicted in Fig. \ref{figqdlagen}. The angle of the light path with respect to the QD plane is only by way of illustration. Hereinafter, the electromagnetic wave is assumed to propagate perpendicular to the QD plane. The total light reflection due to the QD structure is a superposition of the reflections of all interfaces. Taking into account all interference terms between the individual QD layers and surface reflections, the total sample reflection is calculated analytically in analogy with Ref. \cite{tas92}, and can be described as
\begin{equation}
|r_{tot}|^2=|r_s + \sum_k C_k r_{QD}e^{i\phi_k} + C_b r_b e^{i\phi_b} +\sum_k D_k r_s r_{QD}^2 e^{2i\phi_k} +\cdots +M|^2.\label{eq1h2}
\end{equation}
Where $r_s$ denotes the surface reflectivity $r_s=\frac{n_{GaAs}-n_{air}}{n_{GaAs}+n_{air}}$, with $n_{GaAs}$ the refractive index of GaAs. $r_{QD}$ and $r_b$ represent the reflectivity of a single QD layer and the back surface, respectively. The summation takes into account the contributions of the individual QD layers with a relative phase $\phi_k$ induced by the optical path length of the GaAs barrier layers. $C$ and $D$ are correction terms which take into account the amplitude change due to reflection and absorption of the different layers, and $M$ represents the higher order and mixed reflection terms. That are the terms due to multiple reflections on different QD layers. The first three terms on the righthand side of Eq. (\ref{eq1h2}) are the zeroth order reflections (single reflection) whereas the double reflection on the QD layers is taken into account by the fourth term, etc. In this model, we will neglect the absorption of the probe light within the GaAs barriers, because the absorption coefficient $\alpha$ of GaAs is small within the energy window of the probe light \cite{bro96,ada05,see02}, and because the GaAs layer thickness $d$ is usually relatively small. Hereby, the amplitude change is primarily determined by the surface of the sample with transmittance $t_s$. In addition, for GaAs the value of the attenuation index is much smaller than the real part of the refractive index ($\kappa \ll n$) within the probe energy window. This means that, we can neglect the induced phase shift: $\delta$ = 0. We can write for the correction terms $D\simeq C = t_s^2=1-r_s^2$, for all $k$.

Within the experiment, discrimination between the surface reflection, including the QDs, and the backside reflection can be made by changing the position of the probe laser focal point. Therefore, we will neglect the contribution of the backside to the total reflection signal. We can rewrite Eq. (\ref{eq1h2}), without loss of generality, as
\begin{equation}
|r_{tot}|^2 = |r_s + r_{QD}\sum_k \frac{C e^{i\phi_k}}{1-Cr_s r_{QD} e^{i\phi_k}}|^2. \label{eq2}
\end{equation}
Due to the fact that $r_{QD}\ll 1$, Eq. (\ref{eq2}) can be simplified by
\begin{equation}
|r_{tot}|^2 \simeq |r_s + Cr_{QD}\sum_k e^{i\phi_k}|^2, \label{eq3}
\end{equation}
hence,
\begin{equation}
|r_{tot}|^2=r_s^2 +C r_s^\ast r_{QD}\sum_k e^{i\phi_k} + C r_s (r_{QD}\sum_k e^{i\phi_k})^\ast +|C r_{QD}\sum_k e^{i\phi_k}|^2.\label{eq4}
\end{equation}
For normal incidence, the reflection coefficient of a thin homogeneous QD layer in the vicinity of its transition energy ($\hbar\omega_0$) can be written as \cite{ste99,and98,ivc95}
\begin{equation}
r_{QD}(\omega)=\frac{-i\Gamma}{(\omega-\omega_0)+i(\Gamma+\gamma)},\label{eq5}
\end{equation}
where $\hbar\Gamma$ denotes the radiative broadening and $\hbar\gamma$ the homogeneous nonradiative broadening of the reflectance spectrum. Rewriting Eq. (\ref{eq5}) as
\begin{equation}
r_{QD}(\omega)= -\frac{(\Gamma +\gamma)+i(\omega-\omega_0)}{(\omega -\omega_0)^2+(\Gamma +\gamma)^2} \Gamma, \label{eq6}
\end{equation}
and introducing the QD reflectivity into Eq. (\ref{eq4}), we obtain
\begin{eqnarray}
|r_{tot}(\omega)|^2&=& r_s^2(\omega) -2C r_s(\omega)[\sum_k \frac{(\Gamma+\gamma)\cos\phi_k +(\omega_0 -\omega)\sin\phi_k}{(\omega -\omega_0)^2+(\Gamma +\gamma)^2}\Gamma] \nonumber \\ &&+C^2 |r_{QD}|^2 \frac{\cos(z\phi_0 -\phi_0)-1}{\cos(\phi_0)-1}.\label{eq7}
\end{eqnarray}
Here, we assume $z$ layers of QDs with equal spacing $d_{QD}$, which introduces a phase difference $\phi_0$ due to the optical path length between adjacent layers. Hereby, the phase of each QD layer becomes $\phi_k$ = $k \phi_0$. The third term in Eq. (\ref{eq7}) is small with respect to the first and second term, and will be neglected hereinafter. Equation (\ref{eq7}) can be simplified to:
\begin{equation}
R_{tot}(\omega)= R_s(\omega)+R_{QD}(\omega).\label{eqrtot}
\end{equation}
The total reflection becomes a summation of the surface reflection $R_s$ and the QD reflection $R_{QD}$. In this case the expression for the QD reflection can be written in the form $R_{QD}$($\omega$) = $\mathcal{L}$($\omega$)$\Gamma$($\omega$), by introducing a line shape factor $\mathcal{L}$($\omega$). Note that we postulate a frequency dependence of $\Gamma$. That is, $\Gamma$ is assumed to be strongly dependent on the QD transition energy.

In two-color pump-probe differential reflection spectroscopy, the photon energy of the excitation pulses is tuned above the GaAs band gap energy. Within the probe energy window which is far below the GaAs band gap, the pump induced changes of the GaAs refractive index can be neglected \cite{sha96,cal01}. Hereby, the refractive index changes are considered to be instantaneous with respect to the carrier occupation dynamics within the QDs. Therefore, the change of the sample dielectric function can be ascribed to the carrier induced change of the QD absorption, $\partial\alpha /\partial\eta$, as a result of energy level occupation in the quantum dot. Here, $\eta$ is defined as the carrier density. In addition, due to the QD energy level occupation the radiative broadening will be altered. Thus for two-color TRDR, the pump-induced reflection change within the probe energy window can be written as
\begin{equation}
\frac{\partial}{\partial \eta} R_{tot}(\omega)=\frac{\partial}{\partial \eta} R_{QD}(\omega)=-\frac{\partial \Gamma}{\partial \eta} \cdot \frac{\partial}{\partial \Gamma} [\mathcal{L}(\omega)\Gamma(\omega)], \label{eq8}
\end{equation}
which can be rewritten to:
\begin{eqnarray}
\frac{\partial}{\partial \eta} R_{tot}(\omega)=\Delta R_{tot}(\omega)&=&-\frac{\partial \Gamma}{\partial \eta} [\Gamma(\omega)\frac{\partial}{\partial \Gamma} \mathcal{L}(\omega) +\mathcal{L}(\omega)\frac{\partial}{\partial \Gamma} \Gamma(\omega)] \nonumber \\
&=&-\Delta \Gamma (\omega) [\Gamma(\omega) \mathcal{L}'(\omega)+\mathcal{L}(\omega)], \label{eq9a}
\end{eqnarray}
with $\Delta \Gamma$ = $\frac{\partial\Gamma}{\partial \eta}$ and $\mathcal{L}'$ = $\frac{\partial}{\partial \Gamma}\mathcal{L}$. Note that the phase $\phi$ in the contribution of the QD layers is defined with respect to the sample surface reflection [Eq. (\ref{eq7})], but can now be interpreted as a relative phase between the individual layers. Moreover, in case of a symmetrical configuration, see Fig. \ref{figqdlagen}, the odd parts, i.e., the terms with $\sin \phi_k$, will cancel. Hence, $\mathcal{L}(\omega)$ becomes a typical Lorentzian line function. Although, we have made some assumptions, Eq. (\ref{eq9a}) is a general expression for the carrier induced reflection change of a plane of identical QDs independent of the sample configuration.

For realistic QD nanostructures, due to the QD-size distribution a summation over all homogeneously broadened QD states $N(\omega_i)$ has to be taken into account. This results in an inhomogeneously broadened QD density of states (DOS) of the ensemble, denoted by $D(\omega)$, such that
\begin{equation}
\sum_i \Delta R_{tot}(\omega_i)\rightarrow -\sum_i [\Delta\Gamma(\omega_i)\mathcal{H}(\omega_i)]N(\omega_i).\label{eq11}
\end{equation}
The terms in brackets can be interpreted as a response function of the QD reflection due to carrier occupation of the QDs, with $\mathcal{H}(\omega)$ = $\Gamma(\omega)\mathcal{L}'(\omega) + \mathcal{L}(\omega)$. Hence, the total reflection change becomes
\begin{equation}
\Delta R_{tot}(\omega) = -\hat{\mathcal{H}}(\omega) D(\omega), \label{eqdrtot}
\end{equation}
with
\begin{equation}
\hat{\mathcal{H}}(\omega) D(\omega)= \int_{-\infty}^{\infty}[\Delta \Gamma \cdot \mathcal{H} \cdot N]d\omega.
\end{equation}
Due to the finite width of $\mathcal{H}$, it is more appropriate to substitute $\sum \Delta \Gamma\mathcal{H}N$ by $\int [\Delta \Gamma\mathcal{H}N]$. From Eq. (\ref{eqdrtot}) we conclude that the differential reflection signal $\frac{\Delta R}{R}$, is determined by the QD DOS and the QD response function $\hat{\mathcal{H}}$, which includes the carrier lifetime $\tau_d=\Gamma^{-1}$, and the carrier induced change of the radiative emission rate $\Delta \Gamma$.

%\begin{acknowledgments}
This work is financially supported by the Dutch Foundation for Research on Matter (FOM).
%\end{acknowledgments}
\newpage
\bibliography{biblio}% Produces the bibliography via BibTeX.

\begin{thebibliography}{19}
\expandafter\ifx\csname natexlab\endcsname\relax\def\natexlab#1{#1}\fi
\expandafter\ifx\csname bibnamefont\endcsname\relax
  \def\bibnamefont#1{#1}\fi
\expandafter\ifx\csname bibfnamefont\endcsname\relax
  \def\bibfnamefont#1{#1}\fi
\expandafter\ifx\csname citenamefont\endcsname\relax
  \def\citenamefont#1{#1}\fi
\expandafter\ifx\csname url\endcsname\relax
  \def\url#1{\texttt{#1}}\fi
\expandafter\ifx\csname urlprefix\endcsname\relax\def\urlprefix{URL }\fi
\providecommand{\bibinfo}[2]{#2}
\providecommand{\eprint}[2][]{\url{#2}}

\bibitem[{\citenamefont{Bogaart et~al.}(2005)\citenamefont{Bogaart, Haverkort,
  Mano, van Lippen, N$\ddot{o}$tzel, and Wolter}}]{bog05a}
\bibinfo{author}{\bibfnamefont{E.~W.} \bibnamefont{Bogaart}},
  \bibinfo{author}{\bibfnamefont{J.~E.~M.} \bibnamefont{Haverkort}},
  \bibinfo{author}{\bibfnamefont{T.}~\bibnamefont{Mano}},
  \bibinfo{author}{\bibfnamefont{T.}~\bibnamefont{van Lippen}},
  \bibinfo{author}{\bibfnamefont{R.}~\bibnamefont{N$\ddot{o}$tzel}},
  \bibnamefont{and} \bibinfo{author}{\bibfnamefont{J.~H.}
  \bibnamefont{Wolter}}, \bibinfo{journal}{Phys. Rev. B}
  \textbf{\bibinfo{volume}{72}}, \bibinfo{pages}{195301}
  (\bibinfo{year}{2005}).

\bibitem[{\citenamefont{Tassone et~al.}(1992)\citenamefont{Tassone, Bassani,
  and Andreani}}]{tas92}
\bibinfo{author}{\bibfnamefont{F.}~\bibnamefont{Tassone}},
  \bibinfo{author}{\bibfnamefont{F.}~\bibnamefont{Bassani}}, \bibnamefont{and}
  \bibinfo{author}{\bibfnamefont{L.~C.} \bibnamefont{Andreani}},
  \bibinfo{journal}{Phys. Rev. B} \textbf{\bibinfo{volume}{45}},
  \bibinfo{pages}{6023} (\bibinfo{year}{1992}).

\bibitem[{\citenamefont{Lakhtakia and Weiglhofer}(1992)}]{lak92}
\bibinfo{author}{\bibfnamefont{A.}~\bibnamefont{Lakhtakia}} \bibnamefont{and}
  \bibinfo{author}{\bibfnamefont{W.~S.} \bibnamefont{Weiglhofer}},
  \bibinfo{journal}{IEE Proc. H} \textbf{\bibinfo{volume}{139}},
  \bibinfo{pages}{217} (\bibinfo{year}{1992}).

\bibitem[{\citenamefont{Slepyan et~al.}(2001)\citenamefont{Slepyan, Maksimenko,
  Kalosha, Hoffmann, and Bimberg}}]{sle01}
\bibinfo{author}{\bibfnamefont{G.~Y.} \bibnamefont{Slepyan}},
  \bibinfo{author}{\bibfnamefont{S.~A.} \bibnamefont{Maksimenko}},
  \bibinfo{author}{\bibfnamefont{V.~P.} \bibnamefont{Kalosha}},
  \bibinfo{author}{\bibfnamefont{A.}~\bibnamefont{Hoffmann}}, \bibnamefont{and}
  \bibinfo{author}{\bibfnamefont{D.}~\bibnamefont{Bimberg}},
  \bibinfo{journal}{Phys. Rev. B} \textbf{\bibinfo{volume}{64}},
  \bibinfo{pages}{125326} (\bibinfo{year}{2001}).

\bibitem[{\citenamefont{Maxwell-Garnett}(1904)}]{max04}
\bibinfo{author}{\bibfnamefont{J.~C.} \bibnamefont{Maxwell-Garnett}},
  \bibinfo{journal}{Philos. Trans. R. Soc. London}
  \textbf{\bibinfo{volume}{203}}, \bibinfo{pages}{385} (\bibinfo{year}{1904}).

\bibitem[{\citenamefont{Lakhtakia and Weiglhofer}(1993)}]{lak93}
\bibinfo{author}{\bibfnamefont{A.}~\bibnamefont{Lakhtakia}} \bibnamefont{and}
  \bibinfo{author}{\bibfnamefont{W.~S.} \bibnamefont{Weiglhofer}},
  \bibinfo{journal}{Acta Cryst. A} \textbf{\bibinfo{volume}{49}},
  \bibinfo{pages}{266} (\bibinfo{year}{1993}).

\bibitem[{\citenamefont{Prasad}(2004)}]{pras04}
\bibinfo{author}{\bibfnamefont{P.~N.} \bibnamefont{Prasad}},
  \emph{\bibinfo{title}{Nanophotonics}} (\bibinfo{publisher}{Wiley and Sons,
  Inc., New Jersey}, \bibinfo{year}{2004}).

\bibitem[{\citenamefont{Aspnes}(1982)}]{asp82}
\bibinfo{author}{\bibfnamefont{D.~E.} \bibnamefont{Aspnes}},
  \bibinfo{journal}{Am. J. Phys.} \textbf{\bibinfo{volume}{50}},
  \bibinfo{pages}{704} (\bibinfo{year}{1982}).

\bibitem[{\citenamefont{Born and Wolf}(1999)}]{bor99}
\bibinfo{author}{\bibfnamefont{M.}~\bibnamefont{Born}} \bibnamefont{and}
  \bibinfo{author}{\bibfnamefont{E.}~\bibnamefont{Wolf}},
  \emph{\bibinfo{title}{Principles of optics}} (\bibinfo{publisher}{Cambridge
  University Press}, \bibinfo{year}{1999}), \bibinfo{edition}{7th} ed.

\bibitem[{\citenamefont{Abel$\grave{e}$s}(1972)}]{abe72}
\bibinfo{author}{\bibfnamefont{F.}~\bibnamefont{Abel$\grave{e}$s}},
  \emph{\bibinfo{title}{Optical propeties of solids}}
  (\bibinfo{publisher}{North-Holland Publishing Company, Amsterdam},
  \bibinfo{year}{1972}).

\bibitem[{\citenamefont{Pedrotti and Pedrotti}(1993)}]{ped93}
\bibinfo{author}{\bibfnamefont{F.~L.} \bibnamefont{Pedrotti}} \bibnamefont{and}
  \bibinfo{author}{\bibfnamefont{L.~S.} \bibnamefont{Pedrotti}},
  \emph{\bibinfo{title}{Introduction to optics}} (\bibinfo{publisher}{Prentice
  Hall, New Jersey}, \bibinfo{year}{1993}), \bibinfo{edition}{2nd} ed.

\bibitem[{\citenamefont{Brozel and Stillman}(1996)}]{bro96}
\bibinfo{author}{\bibfnamefont{M.~R.} \bibnamefont{Brozel}} \bibnamefont{and}
  \bibinfo{author}{\bibfnamefont{G.~E.} \bibnamefont{Stillman}},
  \emph{\bibinfo{title}{Properties of Gallium Arsenide}}
  (\bibinfo{publisher}{Inspec, London}, \bibinfo{year}{1996}),
  \bibinfo{edition}{3rd} ed.

\bibitem[{\citenamefont{Adachi}(2005)}]{ada05}
\bibinfo{author}{\bibfnamefont{S.}~\bibnamefont{Adachi}},
  \emph{\bibinfo{title}{Properties of Group-IV, III-V and II-VI
  semiconductors}} (\bibinfo{publisher}{Wiley and Sons, Inc., Chichester},
  \bibinfo{year}{2005}).

\bibitem[{\citenamefont{Seeger}(2002)}]{see02}
\bibinfo{author}{\bibfnamefont{K.}~\bibnamefont{Seeger}},
  \emph{\bibinfo{title}{Semiconductor physics}} (\bibinfo{publisher}{Springer,
  Berlin}, \bibinfo{year}{2002}).

\bibitem[{\citenamefont{Stefano et~al.}(1999)\citenamefont{Stefano, Savasta,
  and Girlanda}}]{ste99}
\bibinfo{author}{\bibfnamefont{O.~D.} \bibnamefont{Stefano}},
  \bibinfo{author}{\bibfnamefont{S.}~\bibnamefont{Savasta}}, \bibnamefont{and}
  \bibinfo{author}{\bibfnamefont{R.}~\bibnamefont{Girlanda}},
  \bibinfo{journal}{Phys. Rev. A} \textbf{\bibinfo{volume}{60}},
  \bibinfo{pages}{1614} (\bibinfo{year}{1999}).

\bibitem[{\citenamefont{Andreani et~al.}(1998)\citenamefont{Andreani,
  Panzarini, Kavokin, and Vladimirova}}]{and98}
\bibinfo{author}{\bibfnamefont{L.~C.} \bibnamefont{Andreani}},
  \bibinfo{author}{\bibfnamefont{G.}~\bibnamefont{Panzarini}},
  \bibinfo{author}{\bibfnamefont{A.~V.} \bibnamefont{Kavokin}},
  \bibnamefont{and} \bibinfo{author}{\bibfnamefont{M.~R.}
  \bibnamefont{Vladimirova}}, \bibinfo{journal}{Phys. Rev. B}
  \textbf{\bibinfo{volume}{57}}, \bibinfo{pages}{4670} (\bibinfo{year}{1998}).

\bibitem[{\citenamefont{Ivchenko and Pikus}(1995)}]{ivc95}
\bibinfo{author}{\bibfnamefont{E.~L.} \bibnamefont{Ivchenko}} \bibnamefont{and}
  \bibinfo{author}{\bibfnamefont{G.}~\bibnamefont{Pikus}},
  \emph{\bibinfo{title}{Superlattices and other heterostructures: symmetry and
  optical phenomena}} (\bibinfo{publisher}{Springer}, \bibinfo{year}{1995}).

\bibitem[{\citenamefont{Shah}(1996)}]{sha96}
\bibinfo{author}{\bibfnamefont{J.}~\bibnamefont{Shah}},
  \emph{\bibinfo{title}{Ultrafast spectroscopy of semiconductors and
  semiconductor nanostructures}} (\bibinfo{publisher}{Springer},
  \bibinfo{year}{1996}).

\bibitem[{\citenamefont{Callan et~al.}(2001)\citenamefont{Callan, Kim, Roeser,
  and Mazur}}]{cal01}
\bibinfo{author}{\bibfnamefont{J.~P.} \bibnamefont{Callan}},
  \bibinfo{author}{\bibfnamefont{A.~M.~T.} \bibnamefont{Kim}},
  \bibinfo{author}{\bibfnamefont{C.~A.~D.} \bibnamefont{Roeser}},
  \bibnamefont{and} \bibinfo{author}{\bibfnamefont{E.}~\bibnamefont{Mazur}},
  \emph{\bibinfo{title}{Ultrafast dynamics in highly excited GaAs}}, Ultrafast
  physical processes in semiconductors (\bibinfo{publisher}{Academic Press},
  \bibinfo{year}{2001}).

\end{thebibliography}

\end{document}